\documentclass[twocolumn,showpacs,preprintnumbers,amsmath,amssymb]{revtex4}
\usepackage{graphicx}
\usepackage{epsfig}
\usepackage{dcolumn}
\usepackage{bm}
\begin{document}
\title{Chern-Simons like term generation in an extended model of QED under
external conditions}
\author{D. Ebert}
\affiliation{Institut f\"ur Physik, Humboldt--Universit\"at zu Berlin,
D-12489  Berlin, Germany}
\author{V. Ch. Zhukovsky}
\author{A. S. Razumovsky}
\affiliation{Faculty of Physics, Department of Theoretical Physics,
  Lomonosov Moscow State University, 119899 Moscow, Russia}

\date{\today}

\begin{abstract}

The possibility of a Chern-Simons (CS) like term generation in an
extended model of QED, in which a Lorentz and CPT non-covariant
interaction term for fermions is present, has been investigated at
finite temperature and in the presence of a background color
magnetic field. To this end, the photon polarization operator in
an external constant axial-vector field has been considered.
One-loop contributions to its antisymmetric component due to
fermions in the linear order of the axial-vector field have been
obtained. Moreover, the first nontrivial correction to the induced
CS term due to the presence of a weak constant homogeneous color
magnetic field has been derived.
\end{abstract}
 \maketitle
\section*{Introduction}

The Lorentz and CPT invariance of the physical laws have been
confirmed with high accuracy in numerous experiments \cite{1}.
Nevertheless, one may make an assumption that these symmetries,
for some unknown reasons, are only approximate. The modern quantum
field theoretical viewpoint admits the possibility of Lorentz
invariance breaking (and, as a consequence, possible CPT
invariance breaking in the local field theory) through a
spontaneous symmetry breaking mechanism. In other words, even
though the underlying laws of nature have Lorentz and CPT
symmetries, the vacuum solution of the theory could spontaneously
violate these symmetries.

The usual Standard Model does not have dynamics necessary to cause
spontaneous Lorentz and CPT violation. However, the violation
mentioned above could occur in a more complicated theory, i.e.,
the Standard Model Extension (SME)\cite{2}. A basic requirement of
such an extended model is that it preserves fundamental
properties, such as renormalizability, unitarity and gauge
invariance. In contrast to usual electrodynamics with its vacuum
state being invariant under Lorentz and CPT transformations, in
the extended model, this vacuum state  appears to be filled up by
``fields'',
which have
a certain orientation in space, and this is the cause of Lorentz
symmetry breaking. Technically, a realization of this violation
might be obtained through adding two different kinds of CPT-odd
interaction terms.
The first of them represents a four dimensional analogue of the
well known Chern-Simons term
$\frac{1}{2}\eta_{\mu}\varepsilon^{\mu\alpha\beta\gamma}F_{\alpha\beta}A_{\gamma}$
with a constant  vector $\eta_{\mu}$, the second one is the
CPT-odd interaction term for fermions
$\overline{\psi}b_{\mu}\gamma^{\mu}\gamma_5\psi$ with a constant
vector $b_{\mu}$ \cite{2}. The latter kind of modification does
not influence the gauge invariance of the action and equations of
motion, but it does modify the dispersion relations for Dirac
spinors \cite{2,3}. The question about the possible dynamical
origin of these constant vectors $\eta_{\mu}$ and $b_{\mu}$
remains an interesting task to be solved. In particular, one of
the possible ways for the Lorentz symmetry to be broken through
the Coleman-Weinberg mechanism \cite{4} was recently suggested for
models, where abelian gauge fields interact with a pseudoscalar
massless Axion field $\theta(x)$. It was shown that in this case,
the vector $\eta_{\mu}$ could be associated with the vacuum
expectation value of the gradient of the Axion field
$\eta_{\mu}\sim\langle\partial_{\mu}\theta\rangle_{0}$ \cite{5}.
At the same time, the pseudovector field $b_{\mu}$  might be
related to some constant background torsion in the large scale
Universe,
$b_{\mu}\sim\varepsilon_{\mu\nu\lambda\delta}T^{\nu\lambda\delta}$
\cite{6}. Moreover, such a CPT-odd term could be
generated by chiral fermions \cite{7}. A modification of QED
resulting in the appearance of a CS like term may predict the
phenomena known as birefringence of light \cite{Rol, 2}.
As it was mentioned above, a CPT-odd interaction term for fermions is
also possible in the framework of the SME, and, in this
case, there arises a natural question about a possibility of
generation of the CS like term through radiative corrections
from the fermionic
sector of the general theory.


There are many papers devoted to investigating such an interesting
possibility, when a constant pseudovector field is present in the
theory (see, e.g., \cite{9}---\cite{13}).
It was shown, that the presence of the background  vector
$b_{\mu}$, indeed, leads to the radiatively induced Chern-Simons
term, i.e., to the modified value of the classical tree level
vector  $\eta_{\mu}$. However, there was an ambiguity in the
definition of this correction, which was supposed to be due to the
choice of the regularization procedure \cite{9}. But, as it has
been clearly shown in one of the recent papers \cite{14}, the
magnitude  of this effect does not depend on the regularization
scheme, but only on the requirement that the maximal residual
symmetry, being the small group of the specific vector $b_{\mu}$,
is realized at quantum level order-by-order in the perturbation
theory. This leads to a unique and non-vanishing value of the
radiatively induced CS coefficient. Yet another question, which
also seems very interesting for investigation, is the temperature
dependence of this generated term. In the present paper, we study
the one-loop contributions to the antisymmetric component of the
photon polarization operator in an external constant axial-vector
field $b_{\mu}$ at finite temperature. These contributions, due to
fermion loops, are obtained  in the linear order in the pseudo
vector field. As a result, we obtained the exact analytical
expression for a thermally induced Chern-Simons term. At the same
time, in considering the  influence of the background axial-vector
field on photon propagation, one should also take into account the
influece of the color vacuum fields on the quark loops. For this
purpose, in the second part of the paper, we calculate in the
one-loop approximation the effective potential for this model,
when both a color magnetic and an axial-vector background field
are present. Then, in the lowest order in the color magnetic
field, we calculate the first nontrivial correction to the result
for the Chern-Simons term obtained earlier \cite{14}.

\section{The Model}
Consider fermions interacting with an electromagnetic field
$A_{\mu}(x)$ and a constant axial-vector field
$b_{\mu}=\rm{const}$. The Lagrangian density of the model is as
follows:
$$ L=L_{\rm{em}}+L_{\rm{Dir}},$$ where
$L_{\rm{em}}=-\frac{1}{4}F_{\mu\nu}F^{\mu\nu}$,
$L_{\rm{Dir}}=\overline{\psi}(\imath\gamma^{\mu}\partial_{\mu}+e\gamma^{\mu}A_{\mu}-m-b_{\mu}\gamma^{\mu}\gamma_5)\psi$.

\begin{figure}
\includegraphics[scale=0.8]{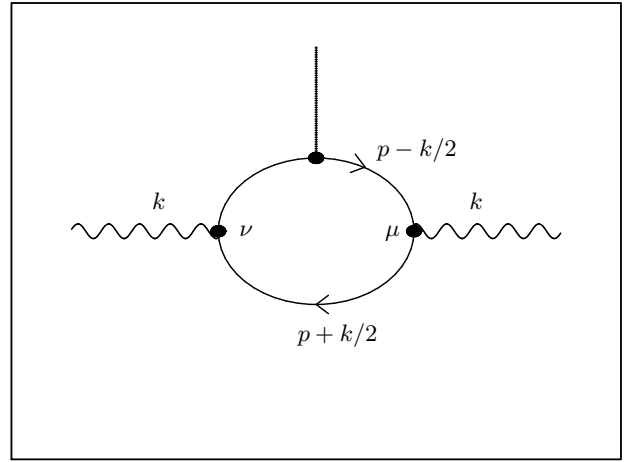}
\begin{picture}(0,0)(0,0)
\put(-95,115){$p-k/2$}\put(-125,45){$p+k/2$}
\put(-92,85){$\mu$}\put(-147,85){$\nu$}\put(-60,95){$k$}
\put(-180,95){$k$}
\end{picture}
\caption{\label{diagr}Photon polarization diagram in a constant
background axial-vector field $b$.}
\end{figure}

Our final objective is to calculate an induced Chern-Simons like
term in the one-loop approximation, and hence, it is sufficient to
calculate the antisymmetric part of the photon polarization
operator
\begin {equation}
\Pi^{\mu\nu}(k)=ie^2\int\frac{d^4p}{(2\pi)^4}{\rm
tr}\left(\gamma^{\mu}S(p+k/2)\gamma^{\nu}S(p-k/2)\right).
\label{1}
\end{equation}
Here, the fermion propagator, modified by the presence of the
axial-vector field $b_{\mu}$, has the form
\begin {equation}
S(p)=\frac{i}{\widehat{p}-m-\widehat{b}\gamma_5}. \label{2}
\end{equation}
This expression can be transformed as follows:
\begin {equation}
S(p)=i\left[\frac{\widehat{p}+m+\widehat{b}\gamma_5}{p^2-m^2+i\varepsilon}-
\frac{2\gamma_5(\widehat{b}m-(bp))(\widehat{p}+m)}{(p^2-m^2+i\varepsilon)^2}\right]+O(b^2),
\label{3}
\end{equation}
where we have retained only the leading terms in the  vector $b$.
Following the remarks made in earlier publications \cite{2}, this
appears to be sufficient to obtain the results needed, i.e., the
antisymmetric part of the polarization operator, given by the
Feynman diagram represented in Fig.~\ref{diagr}.

Introducing the following notations
\begin{eqnarray}
&&A=\frac{m}{p^2-m^2},\,B=\frac{2m(bp)}{(p^2-m^2)^2},\,C^{\mu}=\frac{p^{\mu}}{p^2-m^2},
\cr \cr
&&D^{\mu}=\frac{b^{\mu}}{p^2-m^2}-\frac{2p^{\mu}(bp)}{(p^2-m^2)^2}+\frac{2m^2b^{\mu}}{(p^2-m^2)^2},\nonumber
\\ &&E^{\mu\nu}=-\frac{2mp^{\mu}b^{\nu}}{(p^2-m^2)^2}, \label{4}
\end{eqnarray}
we can rewrite the expression for the propagator (\ref{3}) in the
form
\begin{equation}
S(p)=i(A+B\gamma_5+C^{\mu}\gamma_{\mu}+D^{\mu}\gamma_{\mu}\gamma_5+E^{\mu\nu}\gamma_{\mu}\gamma_{\nu}\gamma_5).
\label{5}
\end{equation}
Our goal is to calculate the antisymmetric part of the
polarization operator $\Pi_{\mu\nu}$ (\ref{1}). Performing trace
operations over spinor indices in (\ref{1}), with the  use of
(\ref{5}), we obtain the required expression  in the leading order
in $b$
\begin{eqnarray}
\Pi_{\mu\nu}^A=\!\!\!\!\!\!&&-4i\varepsilon_{\mu\nu\alpha\beta}\frac{e^2}{(2\pi)^4}\int
d^4p\bigg[(A_1E_2^{\alpha\beta}-A_2E_1^{\alpha\beta})-\nonumber \\
&&-(C_1^{\alpha}D_2^{\beta}-D_1^{\beta}C_2^{\alpha})\bigg],
\label{6}
\end{eqnarray}
where indices 1 and 2 refer to expressions (\ref{4}) for
$A,...,E^{\alpha\beta}$ with $p$ replaced by $p\pm k/2$,
respectively.

\section{Photon polarization operator at finite temperature}
In what follows, calculations at finite temperature will be
performed in the framework of the imaginary time formalism.
Therefore, in order to consider finite temperature, we have to
make the following substitutions
$$\frac{1}{(2\pi)^4}\int d^4p \rightarrow \frac{i}{\beta}\sum^{+\infty}_{n=-\infty}\int \frac{d^3p}{(2\pi)^3},$$
$$p_0\rightarrow
i\omega_0=\frac{i\pi(2n+1)}{\beta},\,n\in\textsl{Z},$$ where
$\omega_0$ is the Matsubara frequency for fermions with
$\beta={\displaystyle\frac{1}{T}}$ as the inverse temperature.
Taking this into account, we rewrite the expression for
$\Pi_{\mu\nu}^{A}$
 (\ref{6}), using (\ref{4}), in the form
\begin{eqnarray}
\Pi_{\mu\nu}^{A,T}=4i\varepsilon_{\mu\nu\alpha\beta}\frac{e^2}{(2\pi)^3}\frac{1}{\beta}\sum^{+\infty}_{n=-\infty}\int^{+\infty}_{-\infty}
d^3p\sum\limits_{i=1}^{3}I_{i}^{\alpha\beta}, \label{7}
\end{eqnarray}
where
$$I_1^{\alpha\beta}=-\frac{k^{\alpha}b^{\beta}}{\Delta_+\Delta_-},$$

$$I_2^{\alpha\beta}=\frac{2k^{\alpha}p^{\beta}\left(2(bp)(p^2+(k/2)^2-m^2)-(bk)(kp)\right)}{\Delta_+^2\Delta_-^2},$$

$$I_3^{\alpha\beta}=-\frac{4m^2k^{\alpha}b^{\beta}(p^2+(k/2)^2-m^2)}{\Delta_+^2\Delta_-^2},$$
and we have introduced the notation
$\Delta_{\pm}\,=\,-[(\vec{p}\pm \vec{k}/2)^2+(\omega_0\pm
k_0/2)^2+m^2]$. In what follows, we discuss only the so called
static limit, when $\vec{k}\rightarrow 0,\,k_0=0$. Other
possibilities of going to the limit $k\rightarrow 0$
 will not be
considered in the present publication. In the static limit,
expression (\ref{7}) takes the form
\begin{widetext}
\begin{equation}
\Pi_{\mu\nu}^{A,T}=4i\varepsilon_{\mu\nu\alpha\beta}k^{\alpha}\frac{e^2}{(2\pi)^3}\frac{1}{\beta}\sum^{+\infty}_{n=-\infty}\int^{+\infty}_{-\infty}
d^3p
\left[-\frac{b^{\beta}(p^2+3m^2)}{\Delta^3}+\frac{4p^{\beta}(bp)}{\Delta^3}\right],
\label{8}
\end{equation}
\end{widetext} where we have taken into account, that
$\Delta_+|_{st.l.}=\Delta_-|_{st.l.}=\Delta=p^2-m^2+i\varepsilon$.
Notice that the vector $b$ is to be time-like $(b^2>0)$, which is
essential for the theory with free fermions interacting with the
axial-vector field. Only in this case, quantization of the  Dirac
field  can be performed in a consistent way \cite{3}. For the sake
of simplicity, though without loss of generality, we choose the
time-like vector in the form $b=(b_0,0,0,0)$, and take $b_0>0$.
Such a restriction does not influence the temperature dependence
of the generated Chern-Simons term, on the one hand, and, on the
other hand, it simplifies all our calculations.

Taking the  above mentioned considerations into account, let us
rewrite (\ref{8}) in spherical coordinates
\begin{widetext}
\begin{equation}
\Pi_{\mu\nu}^{A,T}=2i\varepsilon_{\mu\nu\alpha
0}k^{\alpha}b^0\frac{e^2}{\pi^2}\frac{1}{\beta}\sum^{+\infty}_{n=-\infty}
\int^{\infty}_{0}dpp^2\frac{3\omega_0^2+3m^2-p^2}{(\omega_0^2+p^2+m^2)^3}.
 \label{9}
\end{equation}
\end{widetext}
As it was mentioned in the Introduction, in order to avoid an
ambiguity in definition of radiatively induced CS vector, one
should employ 
the physical 
requirement that the maximal residual symmetry (related to the
small symmetry group of the specific vector $b_{\mu}$) is to be
realized at the quantum level order-by-order in perturbation
theory. Such physical requirement leads to a unique non-vanishing
value of the radiatively induced CS coefficient \cite{14}.
The analysis of the dispersion relations for fermions in an
external axial-vector field demonstrates that there exist
fermions, which would achieve the space-like four-momentum $p^2<0$
at very high energies, a phenomenon which would violate the
Lorentz kinematics in conventional scattering or decay processes.
This means that such electrons interacting with photons would turn
out to be unstable and decay into an electron of the same helicity
and into a pair of electron and positron with opposite helicities.
Therefore, integration over the space momentum in (\ref{9}) should
be restricted by some constant $\Lambda_c$, which
represents a threshold for such a nonphysical reaction. Its value
can be easily calculated using simple kinematic relations
\cite{14}. For pure time-like $b$ it turns out to be equal to
$\Lambda_c={\displaystyle\frac{2m^2}{b_0}}$. Taking this into
account, the integral (\ref{9}) can be written as
\begin{widetext}
\begin{equation}
\Pi_{\mu\nu}^{A,T}=2i\varepsilon_{\mu\nu\alpha0}k^{\alpha}b^0\frac{e^2}{\pi^6}
(\Lambda_c\beta)^3\sum^{+\infty}_{n=-\infty}
{\displaystyle\frac{1}{[(2n+1)^2+(\beta/\pi)^2(\Lambda_c^2+m^2)]^2}}.
\label{10}
\end{equation}
\end{widetext}
The series in the above expression can be easily
summed up \cite{15} to yield

\begin{figure*}[h]
\includegraphics[scale=1.3]{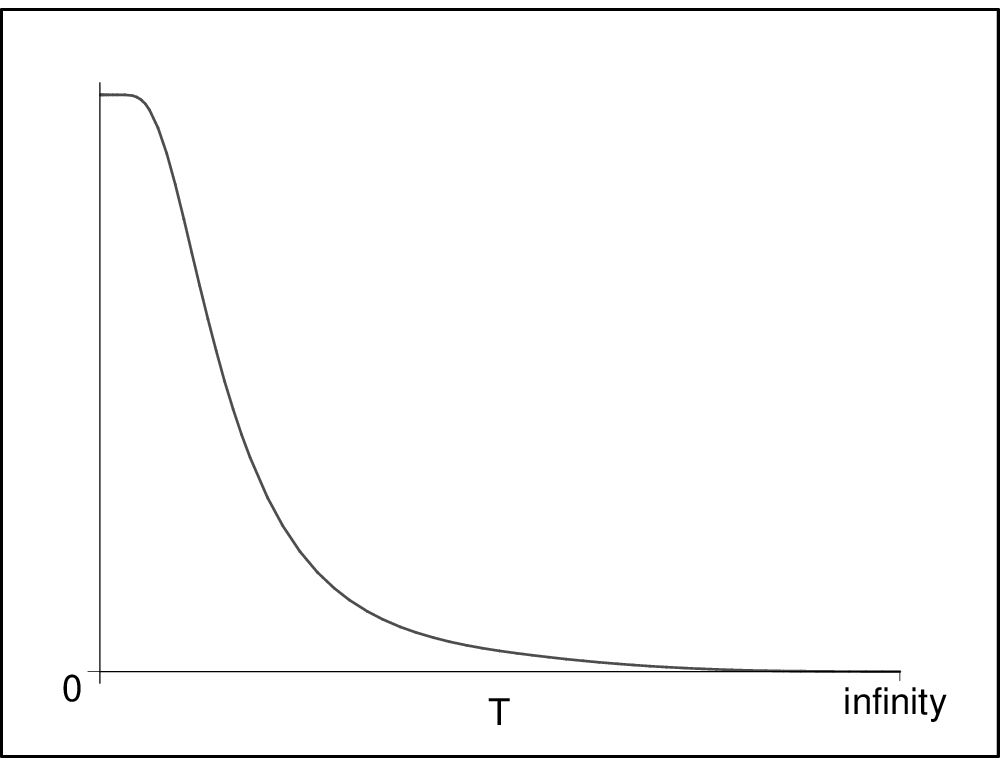}

\begin{picture}(0,0)(0,0)
  \put(-180,220){$\sqrt{\theta_{\mu}^2}$}
\end{picture}

\caption{Temperature dependence of the modulus
$\sqrt{\theta(T)_{\mu}^2}$ of the induced CS vector .}
\label{plot}
\end{figure*}

\begin{widetext}
\begin{equation}
\Pi_{\mu\nu}^{A,T}=i\varepsilon_{\mu\nu\alpha0}k^{\alpha}b^0\frac{e^2}{(2\pi)^2}\left[-\pi
a+\pi a\tanh(\frac{\pi a}{2})^2+2\tanh(\frac{\pi a}{2})\right],
 \label{11}
\end{equation}
\end{widetext}
where the following notation
$a={\displaystyle\frac{\beta\Lambda_c}{\pi}}\sqrt{1+(m/\Lambda_c)^2}\approx{\displaystyle\frac{\beta\Lambda_c}{\pi}}$
was introduced. The curve depicted in Fig.~\ref{plot} represents
(in an arbitrary scale) the modulus $\sqrt{\theta_{\mu}^2}$
 of the radiatively induced thermal Chern-Simons vector $\theta^{\mu}\,=\,(\theta^0,0,0,0)$ with
\begin{equation}
\theta^0(T)\,=\,b^0\frac{e^2}{(2\pi)^2}\left[-\pi a+\pi
a\tanh(\frac{\pi a}{2})^2+2\tanh(\frac{\pi a}{2})\right] \nonumber
\label{cs}
\end{equation}
 as a function of temperature, obtained from expression (\ref{11}). It should be
noticed, that the obtained coefficient has reasonable limiting
values both at $T=0$ and at $T\to\infty$. The first one is
$\Pi_{\mu\nu}^{A}(T=0)=i{\displaystyle\frac{e^2\varepsilon_{\mu\nu\alpha
0}k^{\alpha}b^{0}}{2\pi^2}}$, and it reproduces the result
obtained earlier \cite{14} for the case of vanishing temperature.
At $T\rightarrow\infty$, we have
$\Pi_{\mu\nu}^{A,T}\rightarrow 0$,
which means, that at high temperatures, the Chern-Simons term
generation is completely suppressed and, as a consequence, the
Lorentz and CPT symmetries are completely restored.

\section{The Effective Potential.}

In this section we shall consider the influence of a background
non-abelian gauge field on the quark loop in the above model with
an axial vector field. In order to calculate the effective
potential of the model under these conditions, we shall use the
method based on exact solutions of wave equations that can be
obtained for certain simple configurations of background gauge
fields.
We adopt a  model of quarks in the fundamental representation of
the
 $SU(2)_C$ color group interacting with a non-abelian gauge
field $A_{\mu}=A_{\mu}^aT_a$  and also with an electromagnetic
field
 $F_{\mu\nu}$ and an
axial-vector field $b_{\mu}$. Assuming slow variation of the color
field on the hadronic scale, let us consider, as a first
approximation, a constant non-abelian field $G_{\mu\nu}=$const.
We
 consider the nonabelian background  field to be
rotationally-symmetric (a configuration, which is not possible in
the abelian case)
\begin{eqnarray}
A_a^i=\delta_a^i\sqrt{\lambda},\,\ \!\!\!\!\!\!\!&&A_a^0=0,\,\ G_{ik}^a=g \varepsilon_{ika}\lambda,\nonumber \\
&&\lambda=\rm{const}>0.
 \label{12}
 \end{eqnarray}
We shall assume for later convenience that the following
inequality is valid for the fields introduced
\begin{equation}
b_0^2\ll g^2\lambda \ll m^2,
 \label{12a}
 \end{equation}
although, until special reference to this, we shall not use this
condition. In these fields, the modified Dirac equation looks as
follows:
\begin {equation}
(\gamma^{\mu}\Pi_{\mu}-b_{\mu}\gamma^{\mu}\gamma_5-m)\psi=0,
 \label{13}
\end{equation}
where $\Pi_{\mu}=p_{\mu}-gA_{\mu}^aT_a$. In order to find the
spectrum of stationary states,  it is convenient to consider the
squared Dirac equation
\begin{widetext}
\begin{equation}
\left(\Pi^2-2(\Pi
b)\gamma_5-2\hat{b}\gamma_5\hat{\Pi}-m^2-b^2+\frac{1}{2}g\sigma^{\mu\nu}G_{\mu\nu}^aT_a\right)\psi=0,
 \label{14}
\end{equation}
\end{widetext} or in matrix form
\begin{equation}
(\varepsilon^2-\hat{K})\psi=0, \label{15}
\end{equation}
where operator $\hat{K}$, according to (\ref{12}), has the
following structure
\begin{eqnarray}
\hat{K}=\!\!\!\!\!&&\vec{p}^2+m^2+b_0^2+\frac{3}{4}g^2\lambda-\frac{1}{2}g^2\lambda(\vec{\Sigma}\vec{\tau})-
g\sqrt{\lambda}(\vec{p}\vec{\tau})-\nonumber \\
&&-2b_0\gamma_0\gamma_5(\vec{p}-\frac{1}{2}g\sqrt{\lambda}\vec{\tau})\vec{\gamma},
 \label{16}
\end{eqnarray}
where  $\vec{\Sigma}= \left(
\begin{array}{cc}
  \vec{\sigma} & 0 \\
  0 & \vec{\sigma}
\end{array}
\right)$, and $\vec{\sigma},\,\vec{\tau}$ are Pauli matrices belonging
 to spin and color spaces, respectively.

 Hence, it is simple to obtain the following equation for the
quark spectrum in the  background color field
\begin{eqnarray}
&&\bigg({\vec
p}^2(g^2\lambda+4b_0^2)-a^4+{\displaystyle\frac{1}{4}}g^2\lambda
x^2\bigg)^2-g^2\lambda\bigg(4{\vec p}^2b_0-\nonumber \\
&&-a^2x+{\displaystyle\frac{1}{2}}g\sqrt{\lambda} x^2\bigg)^2=0,
\label{17}
\end{eqnarray}
where we have introduced the notations $a^2={\vec
p}^2+m^2+b_0^2+\frac{3}{4}g^2\lambda-\varepsilon^2$ and
 $x=g\sqrt{\lambda}+b_0$.

Solving this equation, we receive four branches of the spectrum
\begin{equation}
 \left\{
 \begin{array}{ll}
\varepsilon_{1,2}^2=\bigg(|\vec{p}|\pm {\displaystyle\frac{1}{2}}(g\sqrt{\lambda}-2b_0)\bigg)^2+m^2>0,\nonumber \\
\cr
\varepsilon_{3,4}^2=\bigg(\sqrt{{\vec p}^2+g^2\lambda}\pm {\displaystyle\frac{1}{2}}(g\sqrt{\lambda}+2b_0)\bigg)^2+m^2>0.\\
 \end{array}
 \right.
 \label{18}
 \end{equation}
The squared fermion energies are required to be positive as the
necessary condition for the theory to be free from having any
tachyonic modes. With the above values for the spectrum it does
not make any problem now to perform the standard kinematic
considerations \cite{14}, and obtain the value for the cut off
constant from (\ref{18}) using (\ref{12a})
\begin {equation}
 \Lambda_c=\frac{4m^2}{g\sqrt{\lambda}-2b_0}\approx\frac{4m^2}{g\sqrt{\lambda}}\gg
 m.
 \label{19}
\end{equation}
 The one-loop effective action is defined as
\begin {equation}
W_E^{(1)}=\tau\int\frac{dq_4}{2\pi}\sum_{r}\ln{(q_4^2+\varepsilon_r^2)},
 \label{20}
\end{equation}
where $\tau$ is the time interval in euclidian space-time, and
summation over $r$ is assumed to run over all quantum numbers of
quarks, including all spectrum branches, as well as over continuum
of spatial components of the quark momentum. Using the formula
$$\ln{(A/B)}=-\int\limits_0^{\infty}\frac{ds}{s}(\exp{(-sA)}\, - \,\exp{(-sB)})$$
and performing integration over $q_4$, we get for the effective
potential
$V_{\hbox{eff}}^{(1)}=-{\displaystyle\frac{W_E^{(1)}}{\tau L^3}},$
\begin {equation}
V_{\hbox{eff}}^{(1)}=\frac{1}{L^3}\frac{1}{2\sqrt{\pi}}\sum\limits_r\int\limits_0^{\infty}\frac{d
s}{s^{3/2}}\exp{(-s\varepsilon_r^2)}\,-\,{\hbox{c.t.}},
 \label{21}
\end{equation}
where c.t. stands for the counter term  such that
$V_{\hbox{eff}}^{(1)}(b_0=g\sqrt{\lambda}=0)=0$. Taking into
account that
$$\sum\limits_r=\frac{L^3}{(2\pi)^3}\int
d^3p\sum\limits_{n}=\frac{4\pi
L^3}{(2\pi)^3}\int\limits_0^{\infty}p^2dp\sum\limits_n,$$ where
summation $\sum\limits_n$ runs only over the spectrum branches,
and introducing the following notations $z=sm^2$,
$x={\displaystyle\frac{|\vec{p}|}{m}}$,
$\phi^2={\displaystyle\frac{g^2\lambda}{m^2}}$ and
$\psi^2={\displaystyle\frac{b_0^2}{m^2}}$, we rewrite the
effective potential  (\ref{21}) in the form
\begin{widetext}
\begin{eqnarray}
&&\!\!\!V_{\rm
eff}^{(1)}=\frac{m^4}{4\pi^{5/2}}\int\limits_0^{\infty}\frac{dz}{z^{3/2}}\int\limits_0^{\infty}dxx^2
\sum\limits_{\nu=\pm1}e^{-z\left(1+x^2\right)}\times \cr
&&\!\!\!\!\!\!\!\!\!\!\!\!\times\bigg[e^{-z\left(1/4\phi^2-\phi\psi+\psi^2+\nu(\phi-2\psi)x\right)}
+e^{-z\left(5/4\phi^2+\phi\psi+\psi^2+\nu(\phi+2\psi)\sqrt{x^2+\phi^2}\right)}-2\bigg].
 \label{22}
\end{eqnarray}
\end{widetext} The last item in the brackets is the counter term.
To calculate the integral (\ref{22}), we make an expansion of the
integrand in powers of small parameters $\phi, \psi\ll1$. It is
important to mention that, in the general case, the result depends
on the order in which integrations are  performed, i.e.,
$$\int\limits_0^{\infty}{\displaystyle\frac{dz}{z^{3/2}}}e^{-z}\int\limits_0^{\infty}dxx^2e^{-zx^2}f(z,x)$$
or
$$\int\limits_0^{\infty}dxx^2\int\limits_0^{\infty}{\displaystyle\frac{dz}{z^{3/2}}}e^{-z(1+x^2)}f(z,x),$$
where $f(z,x)$ is the integrand  after expansion. The reason for
this is that both expressions, generally speaking, are divergent.
This ambiguity can be eliminated when we apply a certain
regularization procedure, for instance the physical cut off
regularization. This means the integration over
$x={\displaystyle\frac{|\vec{p}|}{m}}$ should be limited  from
above by the cut off (\ref{19})
$$M=\Lambda_c/m={\displaystyle\frac{4m}{g\sqrt{\lambda}}}.$$
Further calculations are made with the help of the relation
$$\int\limits_0^{\infty}\frac{dz}{z^{3/2}}e^{-z(1+x^2)}z^n=(1+x^2)^{(\frac{1}{2}-n)}\Gamma(n-\frac{1}{2}).$$

Thus, eventually, for the one-loop effective potential, we obtain
\begin {equation}
V_{\hbox{eff}}^{(1)}=\frac{m^4}{4\pi^2}\left(I_{\phi}+I_{\psi}+I_{\phi\psi}\right),
 \label{23}
\end{equation}
where
\\
\begin{widetext}
\begin{eqnarray}
\label{24}
\!\!\!\!\!\!\!\!\!\!\!\!\!\!\!\!\!\!\!\!\!\!\!\!\!\!\!\!\!\!\!\!\!\!\!\!\!\!\!\!\!\!\!\!\!I_{\phi}=\!\!\!\!\!\!
&&-\frac{M^3}{\sqrt{1+M^2}}\phi^2+\frac{1}{48}\Bigg(24\ln{(M+\sqrt{M^2+1})}-\frac{24M+35M^3+14M^5}{(1+M^2)^{5/2}}\Bigg)\phi^4-
\nonumber \\ \nonumber \\
&&-\frac{1}{384}M^3\Bigg(\frac{183+408M^2+312M^4+80M^6}{(1+M^2)^{9/2}}\Bigg)\phi^6+O\left(\phi^8\right),
\end{eqnarray}
\begin{eqnarray}
\!\!\!\!\!\!\!\!\!\!\!\!I_{\psi}=&&\!\!\!\!\!\!\!
4\left(\frac{M}{\sqrt{1+M^2}}
-\ln{(M+\sqrt{M^2+1})}\right)\psi^2+\frac{1}{3}M^3\frac{1-2M^2}{(1+M^2)^{5/2}}\psi^4+O\left(\psi^6\right),
\nonumber
\end{eqnarray}
\begin{eqnarray}
\!\!\!\!\!\!\!\!\!\!\!\!\!\!\!\!\!\!\!\!\!\!\!\!\!\!\!\!\!\!\!\!\!\!\!\!\!\!\!\!\!\!\!\!\!\!\!\!\!\!\!\!\!\!\!\!
I_{\psi\phi}=
\frac{M^3}{(1+M^2)^{3/2}}\psi\phi^3-\frac{3}{2}\frac{M^3}{(1+M^2)^{5/2}}\psi^2\phi^2+O\left(\phi^3\psi^3\right).
\nonumber
\end{eqnarray}
\end{widetext}

\begin{figure*}[h]
\includegraphics[scale=1.3]{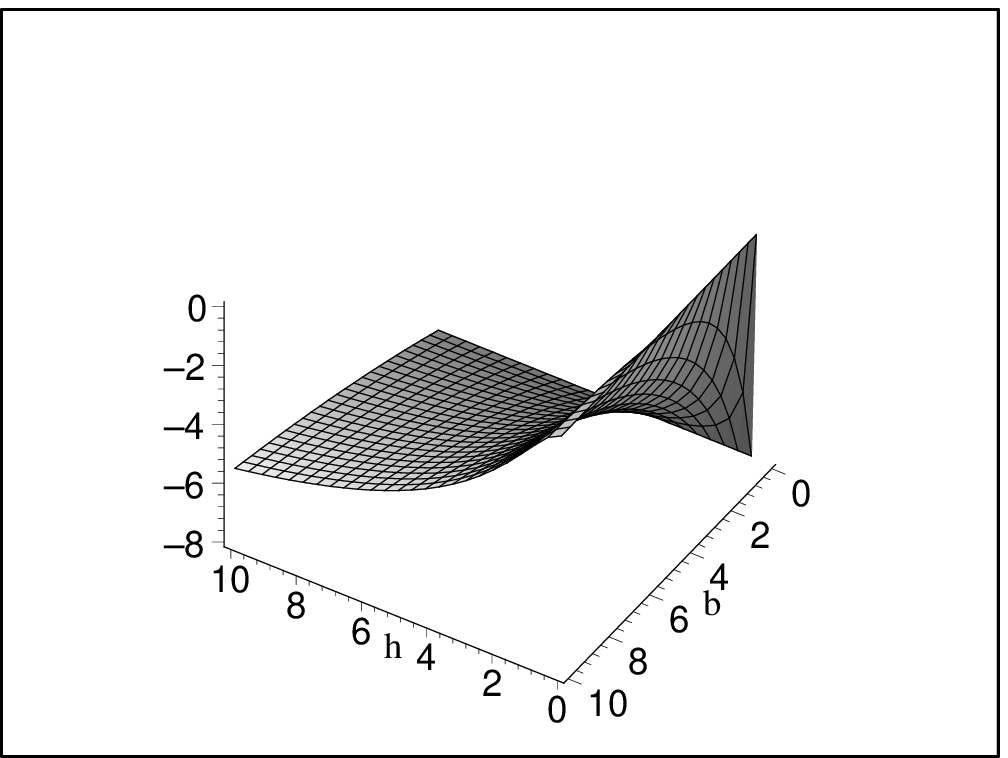}

\begin{picture}(0,0)(0,0)
\put(-140,170){$V_{\hbox{eff}}^{(1)}$}
\end{picture}

\caption{The effective potential $V_{\hbox{eff}}^{(1)}$ as a
function of chromomagnetic and axial-vector field dimensionless
parameters $h={\displaystyle\frac{g\sqrt{\lambda}}{m}}\times
10^{-2}$ and
 $b={\displaystyle\frac{b_0}{m}}\times 10^{-3}$.}
 \label{potent}
\end{figure*}

The  plot of the effective potential $V_{\hbox{eff}}^{(1)}$ as a
function of chromomagnetic and axial-vector fields is depicted in
Fig.~\ref{potent}, where the effective potential is measured in
units of ${\displaystyle\frac{m^4}{4\pi^2}}$, and dimensionless
parameters $h$ and $b$  for the color field and axial-vector field
are defined as $h={\displaystyle\frac{g\sqrt{\lambda}}{m}}\times
10^{-2} $ and $b={\displaystyle\frac{b_0}{m}}\times 10^{-3}$,
respectively. The analysis of this plot demonstrates that with
increasing strength of the color field the contribution of the
axial-vector component decreases.
Let us address to the part of (\ref{24}), which corresponds to the
pure chromomagnetic field contribution $I_{\phi}$. Despite the
presence of  terms of the order of $O(\phi^2)$ and $O(\phi^4)$ in
the expression, the real contribution of the color field to the
effective potential is provided only by the  term of the order
$O(\phi^6)$. This happens because the formal limit of the first
term $-M^2\phi^2$ at $M\rightarrow\infty$ is a pure number, and
its contribution disappears, when one considers the limit of the
whole expression (\ref{23}) (this corresponds to the point
$(h,b)=(0,0)$ in Fig. 3, where the effective potential vanishes,
according to our choice of the counter term). As for the  term
$O(\phi^4)$, its leading  contribution with $M \to \infty$ has the
form
${\displaystyle\frac{1}{8\pi^2}}\ln{(M)}m^4\phi^4\sim\ln{({\displaystyle\frac{m^2}{gH}})}(gH)^2$,
 where  $H=\sqrt{3}g\lambda$ is the field strength,
defining, thereby, the renormalized values of the field and the
charge. Thus, the first  nontrivial finite contribution looks as
$I_{\phi^6}=-{\displaystyle\frac{5}{24}}\bigg({\displaystyle\frac{gH}{m^2\sqrt{3}}}\bigg)^3$,
which exactly coincides with the result obtained in \cite{17}
(actually, this is true  for the term $O(\phi^8)$ as well).
This result  differs from the case of the abelian fields, where
the lowest  term in the expansion of the one-loop effective
potential is of the order of the fourth power of the field
strength $O(gH)^4$, whereas, in the nonabelian theory, besides
quadratic parameters like $F_{\mu\nu}^2$ and
$F_{\mu\nu}\tilde{F^{\mu\nu}}$, new invariant parameters, such as
the cubic one  $F_{\mu\nu}^aF^{\nu
b}_{\lambda}F^{\lambda\mu}_{c}\varepsilon_{abc}$ are possible, and
they may form the lowest correction of the type obtained above.

\section{Radiative correction to the CS coefficient in the presence of
  a  weak color magnetic field}

In this section, we shall demonstrate how the presence of a color
magnetic field corrects the value of the induced Chern-Simons
vector \cite{14}. To this end, we calculate the antisymmetric part
of the polarization operator, when  both chromomagnetic and
axial-vector fields are present in the background. The modified
fermion propagator, in this case, takes the form
\begin {equation}
S(p)=\frac{i}{\hat{\Pi}-m-\hat{b}\gamma_5},
 \label{25}
\end{equation}
where, as before, $\Pi_{\mu}=p_{\mu}-gA_{\mu}^aT_a$. This
expression may be rewritten as
\begin {equation}
S(p)=i{\displaystyle\frac{\hat{\Pi}+m+\hat{b}\gamma_5}{\Pi^2+2(\Pi
b)\gamma_5-2m\hat{b}\gamma_5-m^2+b^2+\frac{1}{2}g(\sigma G)}}.
 \label{26}
\end{equation}

Taking the following relation into consideration
$\left[\gamma_5,\sigma_{ik}\right]=\left[\gamma_0\gamma_5,\sigma_{ik}\right]=0$
for $i,k=\overline{1,3}$, one can perform, for the background
field configuration (\ref{12}) admitted, a correct expansion of
the propagator in  powers of $b$ restricting oneself, as before,
to the term linear in $b_0$. This results in
\begin{widetext}
\begin {equation}
S(p)=i\left(\hat{\Pi}+m+\hat{b}\gamma_5\right)\left[{\displaystyle\frac{1}{\Pi^2-m^2+1/2g(\sigma
G)}}-{\displaystyle\frac{2(\Pi
b)\gamma_5-2m\hat{b}\gamma_5}{\left(\Pi^2-m^2+1/2g(\sigma
G)\right)^2}}\right].
 \label{27}
\end{equation}
\end{widetext}

Further, taking into account, that the antisymmetric part of PO
appears as a structure proportional to an antisymmetric tensor, we
expand (\ref{27}) in  powers of $(\sigma G)$ and keep only the
linear term. Then,
\begin{widetext}
\begin{eqnarray}
S(p)\!\!\!\!\!\!\!&&=i\bigg[A+B\gamma_5+C_{\mu}\gamma^{\mu}+D_{\mu
ik}\gamma^{\mu}\sigma^{ik}+E_{ik}\sigma^{ik}+F_{ik}\gamma_0\gamma_5\sigma^{ik}+
I\gamma_0\gamma_5+J_{\mu}\gamma_{\mu}\gamma_5+K_{\mu}\gamma^{\mu}\gamma_0\gamma_5+\nonumber
\\
&&+L_{\mu ik}\gamma^{\mu}\sigma^{ik}\gamma_5+M_{\mu
ik}\gamma^{\mu}\sigma^{ik}\gamma_0\gamma_5+N_{ik}\sigma^{ik}\gamma_5
\bigg],
 \label{28}
\end{eqnarray}
\end{widetext}
where we have introduced a new notations for
\begin{widetext}
\begin{eqnarray}
&&\!\!\!\!\!\!\!\!A={\displaystyle\frac{m}{\Delta}},\,\,
B=-\frac{2mb_0p_0}{\Delta^2},\,\,
C_{\mu}=\frac{\Pi_{\mu}}{\Delta},\,\, D_{\mu ik
}=-\frac{\Pi_{\mu}gG_{ik}}{2\Delta^2},\,\,
E_{ik}=-\frac{mgG_{ik}}{2\Delta^2},\,\, F_{ik}=-\frac{b_0
gG_{ik}}{2\Delta^2}\bigg(1+\frac{2m^2}{\Delta}\bigg),\nonumber
\\ \nonumber \\
&&\!\!\!\!\!\!\!\!I=\frac{b_0}{\Delta}\bigg(1+\frac{2m^2}{\Delta}\bigg),\,\,
J_{\mu}=-\frac{2p_0b_0\Pi_{\mu}}{\Delta^2},\,\,
K_{\mu}=\frac{2mb_0\Pi_{\mu}}{\Delta^2},\,\, L_{\mu
ik}=\frac{2p_0b_0\Pi_{\mu}gG_{ik}}{\Delta^3},\,\, M_{\mu ik}=
-\frac{2m\Pi_{\mu}gG_{ik}b_0}{\Delta^3},\nonumber \\ \nonumber \\
&&\!\!\!\!\!\!\!\!N_{ik}=\frac{2mp_0b_0gG_{ik}}{\Delta^3}
 \label{29}
 \end{eqnarray}
 \end{widetext}
 with $\Delta=\Pi^2-m^2$.

The polarization operator is defined as before (see (\ref{1})),
where the trace operation now should be perform over color indices
as well.
In order to obtain the antisymmetric part of PO, we have to
calculate  the trace over spinor indices. Excluding from the
resulting expression the terms that refer to the pure color
magnetic field, we obtain
\begin{widetext}
\begin{eqnarray}
 \label{31}
\Pi^A_{\rho\sigma}\!\!\!\!\!\!&&=-ie^2{\rm
tr}_c\int\frac{d^4p}{(2\pi)^4}\Bigg[4i\varepsilon_{\rho\sigma\mu0}(K_{1}^{\mu}A_2-K_{2}^{\mu}A_1)
+4(g_{\rho\delta}g_{\sigma\mu}-g_{\rho\mu}g_{\sigma\delta})\varepsilon_{ik0\delta}(M_{1}^{\mu
ik}A_2-A_1M_{2}^{\mu ik})+ \nonumber
\\ \cr
  &&+4\varepsilon_{\rho\sigma
ik}(A_1N_2^{ik}-N_1^{ik}A_2)+4i\varepsilon_{\rho\sigma\mu0}(C_1^{\mu}I_2
-I_1C_2^{\mu})+4i\varepsilon_{\rho\sigma\mu\nu}(C_1^{\mu}J_2^{\nu}-C_2^{\mu}J_1^{\nu})
+4\varepsilon_{\rho\sigma\mu\delta}(g_{i\nu}g_{k\delta}-\nonumber
\\ \cr
&&-g_{k\nu}g_{i\delta})(C_1^{\mu}L_2^{\nu ik}-L_1^{\nu
ik}C_2^{\mu})-4i\varepsilon_{ik\mu\delta}\varepsilon_{lm0\alpha}\varepsilon_{\rho\sigma\alpha\delta}(D_1^{\mu
ik}F_2^{lm}-F_1^{lm}D_2^{\mu
ik})+4\varepsilon_{\rho\sigma\delta0}(g_{i\mu}g_{k\delta}-g_{k\mu}g_{i\delta})\times
\nonumber
\\ \nonumber \\
&&\times(D_1^{\mu ik}I_2-I_1D_2^{\mu
ik})+4\varepsilon_{\rho\sigma\delta\nu}(g_{i\mu}g_{k\delta}-g_{k\mu}g_{i\delta})(D_1^{\mu
ik}J_2^{\nu}-J_1^{\nu}D_2^{\mu
ik})+4i\varepsilon_{\rho\sigma\alpha\delta}\bigg((g_{i\mu}g_{k\delta}-g_{k\mu}g_{i\delta})\times
\nonumber \\ \nonumber \\
&&\times(g_{l\nu}g_{m\alpha}-g_{m\nu}g_{l\alpha})-
\varepsilon_{ik\mu\delta}\varepsilon_{lm\nu\alpha}\bigg)(D_1^{\mu
ik}L_2^{\nu lm}-L_1^{\nu lm}D_2^{\mu
ik})+E_1^{ik}K_2^{\mu}\bigg(4\varepsilon_{ik\rho\delta}(g_{\delta\sigma}g_{\mu0}-g_{\delta\mu}g_{\sigma0}+g_{\delta0}g_{\sigma\mu})-
\nonumber
\\ \cr
&&-4\varepsilon_{\delta\sigma\mu0}(g_{i\rho}g_{k\delta}-g_{k\rho}g_{i\delta})\bigg)+E_2^{ik}K_1^{\mu}\bigg(\rho\leftrightarrow\sigma\bigg)
+E_1^{ik}M_2^{\nu
lm}\bigg(4i\varepsilon_{ik\rho\delta}\varepsilon_{lm0\alpha}\varepsilon_{\delta\alpha\sigma\nu}+4i\varepsilon_{lm0\alpha}\times\nonumber
\\ \cr
&&\times(g_{i\rho}g_{k\delta}-g_{k\rho}g_{i\delta})(g_{\delta\sigma}g_{\nu\alpha}-g_{\delta\nu}g_{\sigma\alpha}+g_{\delta\alpha}g_{\sigma\nu})
\bigg)+E_2^{ik}M_1^{\nu
lm}\bigg(\rho\leftrightarrow\sigma\bigg)\Bigg],
\end{eqnarray}
\end{widetext}
where indices 1 and 2  mean that we use (\ref{28})
for $A,...,N^{ik}$, with $p$ replaced by $p\pm k/2$ respectively,
 symbol $(\rho\leftrightarrow\sigma)$ is used to denote the same
expression as the previous one up to permutation of $\rho$ and
$\sigma$, and  ${\rm tr}_c$ stands for the trace over color
indices.

Each integral in (\ref{31}) has the general structure, which may
be represented in the form
$$\int\frac{d^4p}{(2\pi)^4}\frac{\Phi_{\rho\sigma}(p,gA^aT_a,m)}{(\Delta_1\Delta_2)^n}.$$
Here, $\Delta_{1,2}=\Pi_{1,2}^2-m^2$, and $n=2,3$ for different
integrals. It should be mentioned that in the static limit (the
rest frame of reference) $(\vec{k}\rightarrow 0,\,k_0=0)$, the
denominator is equal to
$(\Delta_1\Delta_2)^n=(p_0^2-(\vec{p}-g\vec{A}_a{\displaystyle\frac{\tau^a}{2}})^2-m^2)^{2n}.$
Thus, expanding the integrand up to terms of the order
$O(gA\tau)^4$ and calculating the trace over color indices, we
integrate over $p$ with the upper limit equal to the constant
$\Lambda_c$, as it was prescribed in the previous section.  As a
result, we obtain for the antisymmetric part of PO
\begin {equation}
\Pi_{\rho\sigma}^A=-ie^2\frac{2}{\pi^2}\varepsilon_{\rho\sigma\mu0}k_{\mu}b_0
\Bigg[-\frac{1}{2}+\frac{15}{32}\bigg(\frac{g\sqrt{\lambda}}{m}\bigg)^2+O\bigg(\frac{g\sqrt{\lambda}}{m}\bigg)^4\Bigg].
 \label{34}
\end{equation}

We remind that the contribution of a pure color magnetic field to
the antisymmetric part of PO is given by the formula \cite{16}
\begin {equation}
\Pi_{\rho\sigma}^A(b_0=0,\lambda\neq
0)=ie^2\frac{5}{24\pi^2}\varepsilon_{\rho\sigma\mu}k^{\mu}\frac{g^3\lambda^{3/2}}{m^2}.
\label{35}
\end {equation}

The first term in the brackets
of our result (\ref{34}) refers to the induced Chern-Simons term
$\Pi^A(b_0\neq~0,\lambda=0)$,  when only the axial-vector field is
present in the theory, and this is in complete agreement with the
result, obtained in \cite{14}. The second term gives the
correction, calculated with both fields present
$\triangle\Pi^A(b_0\neq0,\lambda\neq0)$, and its value depends on
their relative strength
$$\frac{\triangle\Pi^A(b_0\neq0,\lambda\neq0)}{\Pi^A(b_0=0,\lambda\neq0)}\sim\frac{b_0}{g\sqrt{\lambda}}\ll1.$$

At the same time, the ratio of the contributions induced by the
axial vector and the color fields separately
$$\frac{\Pi^A(b_0\neq0,\lambda=0)}{\Pi^A(b_0=0,\lambda\neq0)}\sim\frac{b_0}{g\sqrt{\lambda}}\bigg(\frac{m}{g\sqrt{\lambda}}\bigg)^2$$
substantially depends not only on the relation of the fields,  but
also on the ratio of the fermionic mass and the strength of the
color field. Therefore, under the condition (\ref{12a}), when this
ratio is large, the color field and the axial vector field may
provide comparable contributions to the induced  CS vector.

\section*{Conclusions}
We have calculated the one-loop fermion contribution to the
antisymmetric part of the photon polarization operator in an
external constant axial-vector field $b_{\mu}$. The result was
obtained in the linear order in the pseudo vector field, using the
physical cut off regularization scheme. The analysis of the
temperature dependence of the obtained expression allows us to
conclude that generation of a  Lorentz- and CPT-odd term may occur
at any physical value of temperature.
In particular, we have reproduced the standard result for the case
of vanishing temperature, $T=0$, \cite{14}. Moreover, we have
shown that this effect is completely suppressed in the limit of
very high temperature, $T \to\infty$, when the theory restores its
Lorentz and CPT symmetries.

The influence of the vacuum field, modelled by a constant
nonabelian color magnetic field, on  generation of a Chern-Simons
term has been considered in the one-loop approximation. We have
constructed the effective potential for this model with
consideration
of both the axial-vector field and a nonabelian color field. We
have  demonstrated that with increasing strength of the color
field the contribution of the  axial-vector component to the
effective potential decreases. The first nontrivial correction to
the induced topological
CS vector
 due to the
presence of a weak (with respect to fermion mass) color magnetic
field has been obtained and its relative contribution to the total
CS coefficient has been estimated.

It is important to notice that the possible presence of an
antisymmetric part of the photon polarization operator
demonstrates  spatial anisotropy. This may provide one of the
physical mechanisms for possible unusual phenomena in the
propagation of light through the universe, i.e.,
rotation of the plane of polarization of electromagnetic radiation
propagating over cosmological distances (the effect, different
from the usual Faraday rotation, which was
 discussed in recent publications \cite{2}).

In the present work, as in the series of papers mentioned in the
Introduction, we have used the extended model of QED, where the
Lorentz and CPT non-covariant interaction term for fermions (a
constant axial-vector field) is present. Interactions of photons
with fermion loops in this background field lead to  the phenomena
mentioned above. However, it should be mentioned that  the
dynamical origin of this pseudovector field, in spite of numerous
efforts, still remains to be explained \cite{5,6,7}.
\section*{Acknowledgements}
We would like to thank  Prof.M.Mueller-Preussker for discussions
and hospitality
at the HU-Berlin.   One of the authors (A. R.) acknowledges
financial support by the Leonhard Euler program of the German
Academic Exchange Service (DAAD) extended to him, while part of
this work was carried out.
The other author (V.Ch.Zh.) acknowledges support by DAAD and
partly by the DFG-Graduiertenkolleg ``Standard Model''.


\begin{thebibliography}{20}
\bibitem{1}
 K. Hagiwara et al., Phys. Rev. D {\bf 66}, 010001 (2002).
 \bibitem{Rol}
 B. Nodland and J. P. Ralston, Phys. Rev. Lett. {\bf 78}, 3047
 (1997), astro-ph/9704196; B. Nodland and J. P. Ralston, Phys. Rev. Lett. {\bf 79},
 1958 (1998), astro-ph/9708114.
\bibitem{2}
S. M. Carroll, G. B. Field and R. Jackiw, Phys. Rev. D {\bf41},
1231 (1990); D. Colladay and V. A. Kostelecky, Phys. Rev. D
{\bf55}, 6760 (1997), hep-ph/9703464; D. Colladay and V. A.
Kostelecky, Phys. Rev. D {\bf58}, 11602 (1998), hep-ph/9809521; S.
R. Coleman and S. L. Glashow, Phys. Rev. D {\bf 59}, 116008
(1999), hep-ph/9812418.

\bibitem{3}
V. A. Kostelecky and R. Lehnert, Phys. Rev. D {\bf 63}, 065008
(2001), hep-th/0012060.
\bibitem{4}
S. R. Coleman and E. Weinberg, Phys. Rev. D {\bf D7}, 1888 (1973).
\bibitem{5}
A. A. Andrianov, R. Soldati and L. Sorbo, Phys. Rev. D {\bf 59},
025002 (1999), hep-th/9806220.
\bibitem{6}
I. L. Shapiro, Phys. Rept. {\bf 357}, 113 (2001), hep-th/0103093.
\bibitem{7}
G. E. Volovik, Sov. Phys. JETP Lett. {\bf 70}, 1 (1999),
hep-th/9905008; G. E. Volovik and A. Vilenkin, Phys. Rev. D {\bf
62}, 025014 (2000), hep-ph/9905460.

\bibitem{9}
R. Jackiw and V. A. Kostelecky, Phys. Rev. Lett. {\bf 82}, 3572
(1999), hep-ph/9901358.

\bibitem{10}
M. Perez-Viktoria, Phys. Rev. Lett. {\bf 83}, 2518 (1999),
hep-th/9905061; M. Perez-Viktoria, J. High Energy Phys. {\bf 04},
032 (2001).

\bibitem{11}
M. Chaichian, W. F. Chen and R. Gonzalez Felipe, Phys.Lett. B {\bf
503}, 215 (2001), hep-th/0010129.

\bibitem{12}
J. M. Chung and P. Oh, Phys. Rev. D {\bf 60}, 067702 (1999),
hep-th/9812132; J. M. Chung, Phys. Rev. D {\bf 60}, 127901 (1999),
hep-th/9904037; J. M. Chung and B. K. Chung, Phys. Rev. D {\bf
63}, 105015 (2001), hep-th/0101097.
\bibitem{13}
W. F. Chen, Phys.Rev. D {\bf 60}, 085007 (1999),  hep-th/9903258.
\bibitem{14}
A. A. Andrianov, P. Giacconi and R. Soldati, jhep/022002030.
\bibitem{15}
A. R. Prudnikov, Yu. A. Brychkov and O. I. Marichev, Integrals and
Series (in Russian), Moscow, 1981.
\bibitem{16}
D. Ebert and V. Ch. Zhukovsky, hep-th/9712016.
\bibitem{17}
I. M. Ternov, V. Ch. Zhukovsky and A. V. Borisov, Quantum
Processes In a Strong External Field (in Russian), Moscow, 1989.
\end{thebibliography}
\end{document}